\documentclass[]{aastex631}

\begin{document}

\title{A 3D Thermophysical Model for Binary Asteroid Systems: Application to the BYORP Effect on (175706) 1996 FG3} 

\author[0000-0002-0786-7307]{Kya C. Sorli}
\affiliation{Department of Astrophysical and Planetary Sciences \\
University of Colorado Boulder, Boulder, CO, USA}
\affiliation{Laboratory for Atmospheric and Space Physics \\
University of Colorado Boulder, Boulder, CO, USA}

\author[0000-0003-4399-0449]{Paul O. Hayne}
\affiliation{Department of Astrophysical and Planetary Sciences \\
University of Colorado Boulder, Boulder, CO, USA}
\affiliation{Laboratory for Atmospheric and Space Physics \\
University of Colorado Boulder, Boulder, CO, USA}

\author[0009-0000-2266-6266]{Rachel H. Cueva}
\affiliation{Smead Department of Aerospace Engineering Sciences \\
University of Colorado, Boulder, CO, USA}

\author[0009-0000-8275-8779]{Chloe J. Long}
\affiliation{Smead Department of Aerospace Engineering Sciences \\
University of Colorado, Boulder, CO, USA}

\author[0000-0002-1847-4795]{Jay W. McMahon}
\affiliation{Smead Department of Aerospace Engineering Sciences \\
University of Colorado, Boulder, CO, USA}

\author[0000-0003-0558-3842]{Daniel J. Scheeres}
\affiliation{Smead Department of Aerospace Engineering Sciences \\
University of Colorado, Boulder, CO, USA}

\begin{abstract}
    Binary asteroids originate from a wide range of evolutionary pathways, and are targets of several previous and upcoming spacecraft missions. Differential heating and radiation on asymmetric asteroids can cause measurable changes in their rotation rates and spin axes, collectively known as the Yarkovsky–O'Keefe–Radzievskii–Paddack (YORP) effect. In binary systems, such radiation-driven torques can cause changes to the mutual asteroid orbits, termed the binary YORP or BYORP effect. To study how binary asteroid shapes and thermophysical properties affect surface temperatures and BYORP, we developed a new 3D thermophysical model. This model can be applied to binary asteroid systems, solitary asteroids, and other airless bodies with complex topography. The model balances direct insolation, 1D conduction, visible light reflection, and mutual heating through scattered infrared radiation. Using 3D ray tracing, we include eclipses, shadowing from horizons and topography, as well as the mutual radiation exchange between the primary and secondary asteroids. Using this model, we perform global temperature modeling of the binary asteroid (175706) 1996 FG3, a target of the Janus mission. At perihelion, we find that the 1996 FG3 system experiences temperatures between $\sim$100 and 475 K. We also find that eclipses and thermal inertia can alter surface temperatures on the secondary by up to 14\%, with a mean difference due to radiation from the primary of just over 1\%. These radiative effects decrease with higher thermal inertia. We also present a model for calculating the BYORP effect using the results of the binary thermophysical model. This model compares well to analytical approximations of the BYORP coefficient $B$, and suggests that thermal effects such as eclipses and thermal inertia can reduce torque in the 1996 FG3 system and alter the BYORP coefficient $B$ by up to several percent. Though small, these second-order effects may produce significant dynamical changes. For 1996 FG3, eclipses alter $B$ by approximately 7\%, resulting in a lower torque on the secondary. In the absence of tidal effects, this change would reduce the contraction of the semimajor axis by about 20 meters over 10,000 years. Mutual radiation from the primary also causes a small nonzero change to $B$, although of an order of magnitude smaller. Our findings suggest that thermal effects can alter temperatures and BYORP calculations sufficiently that they should be included when modeling binaries, and the relative importance of each effect is predicted to vary with the properties of the system being studied. 
\end{abstract}

\section{Introduction and Background}
Thermophysical modeling is an important tool for predicting temperatures and thermal emission of planetary surfaces. For small airless bodies, thermal effects may alter not only surface temperatures and regolith properties, but also orbital evolution. Applications of thermophysical models are varied. They can be used to simulate a body's thermophysical environment, including surface and subsurface temperatures, determine thermal stability of volatiles (e.g., \cite{Hayne2015ThermalTopography, Paige2010DivinerRegion}), interpret infrared (IR) observations to constrain the thermal inertia and regolith properties of the surface (e.g., \cite{Rozitis2020AsteroidEquator}), and predict the magnitude of dynamical effects from solar radiation forces, such as Yarkovsky or YORP \citep{Paddack1969RotationalPressure}.

Roughly 15\% of the Near Earth Asteroid (NEA) populations is estimated to be double asteroids, or binaries  \citep{Pravec2006PhotometricAsteroids, Bottke1996BinaryCraters, Margot2002BinaryPopulation}. Multiple asteroid systems pose increased complexity. An additional body leads to new effects, like eclipses or mutual radiation exchange between bodies (hereafter referred to as "mutual radiation") which could alter temperatures and binary orbital properties. Additionally, multiple systems can provide useful opportunities to learn about other physical and mechanical properties, including masses, densities, and potentially even interior structure \citep{Margot2015AsteroidsBooks}.

Several recent missions have targeted binary asteroids. In 2022, The Double Asteroid Redirect Test (DART) performed a kinetic impact to test asteroid impact hazard mitigation. It struck the secondary of the (65803) Didymos system, known as Dimorphos, and successfully altered its orbital period by about 33 minutes \citep{Thomas2023OrbitalImpact, Cheng2018AIDAObjectives, Cheng2023MomentumDimorphos}. A partnered ESA mission, HERA, will visit the Didymos system to study the physical and thermal properties of the system \citep{Cheng2018AIDAObjectives}. 

The 2023 flyby of the asteroid (152830) Dinkinesh by the Lucy spacecraft revealed it to be a binary \citep{Levison2021LucyGoals, Levison2024ADinkinesh}. The Dinkinesh satellite, Selam, was also found to be a contact binary, likely resulting from a low-velocity encounter between two objects of comparable size. The unexpected discovery of the Dinkinesh secondary emphasizes not only how binaries may be more common than previously believed, but also the complex evolutionary paths they can take. As more missions venture to small bodies (e.g. Hayabusa2 \citep{Shimaki2020ThermophysicalMapping}, OSIRIS-REx \citep{Lauretta2017OSIRIS-REx:Bennu}, and the Emirates Mission to the Asteroids \citep{AlMazmi2023ScienceBelt}), the increasing amount of thermal infrared data will require accurate thermal models for complex asteroid systems, including binary asteroids.

\subsection{Binary Asteroid Dynamics}
Binary asteroids result from multiple dynamical pathways. Most known NEA and main-belt binaries have primaries with relatively rapid rotation rates and comparably smaller synchronously rotating secondaries \citep{Margot2015AsteroidsBooks, Pravec2006PhotometricAsteroids}. A subgroup of binaries have a secondary with asynchronous rotation, such as the former Janus mission target system (35107) 1991 VH \citep{Pravec2016BinaryElongations}. Still others are doubly synchronous, such as the almost equal-mass binary (69230) Hermes \citep{Margot2006Hermes, Jacobson2011LONG-TERMASTEROIDS}.

The dynamical state of a system depends on complex interactions between several factors including asteroid shape, solar radiation pressure and tidal forces. In a solitary rotating body, pressure from solar radiation can induce a time-varying periodic force on the surface known as the Yarkovsky–O’Keefe–Radzievskii–Paddack (YORP) effect \citep{Paddack1969RotationalPressure}. In some cases, YORP can cause sufficient spin-up of a rubble pile asteroid that it breaks apart, potentially forming a binary \citep{Walsh2008RotationalAsteroids}. This is a likely formation mechanism for binary systems \citep{Walsh2015FormationAsteroids}. 

An analog of the YORP effect occurs in binary systems. In binaries, asymmetric solar radiation pressure (SRP) and emission produces a net force or torque that can alter the orbit of the secondary. In the case of a synchronously rotating secondary, this force will be time varying but periodic over the orbital period. This is known as the Binary YORP (BYORP) effect \citep{Cuk2005EffectsNEAs}. BYORP can contract or expand the mutual orbit of an asteroid pair, and may also play an important role in the system's orbital evolution and stability \citep{Goldreich2009TIDALPILES, Jacobson2011LONG-TERMASTEROIDS, McMahon2010SecularBYORP}. Estimates suggest that BYORP may significantly change a binary mutual orbit within as little as $10^{4}$ to $10^{5}$ years \citep{Cuk2005EffectsNEAs}.

To the first order, the evolution of the semimajor axis and eccentricity of a binary mutual orbit is dependent on a single constant term, $B$, known as the BYORP coefficient. It is a normalized 0th-order Fourier coefficient representing the SRP force in the direction parallel to the secondary’s orbital motion \citep{McMahon2010DetailedPopulation}. The BYORP coefficient can be estimated for stable systems with a synchronous secondary.  Current BYORP theory, developed by \cite{McMahon2010SecularBYORP} and based on theory for single-body YORP from \cite{Scheeres2007TheYORP}, uses secular averaging over a heliocentric orbit and models the SRP force as a Fourier series over a secondary mutual orbit to estimate the magnitude of $B$. As $B$ is highly shape dependent, this method requires a known shape model.  

However, this approach assumes a thermal inertia of zero, no eclipsing, no mutual radiation exchange between the bodies, and a system in radiative equilibrium. All of these processes affect surface temperatures and may cause a deviation from the periodic SRP force, especially the drop in temperature from eclipses and the lag in re-emission caused by thermal inertia. This knowledge gap presents an opportunity to use thermophysical modeling to illuminate questions of dynamics and asteroid evolution. 

The potential importance of eclipsing in altering binary evolution has been raised by recent work. \cite{Zhou2024TheAsteroids} found that the component of the binary Yarkovsky effect arising from eclipses can dominate over the radiative force from the primary for certain systems, especially for low-inclination binaries. They also find the binary Yarkovsky effect can be a powerful mechanism for synchronizing secondaries in a prograde orbit, or may cause outward migration for retrograde secondaries. Though these findings only apply to non-synchronous secondaries, they emphasize the importance of understanding how eclipses and other effects may alter BYORP.

\begin{figure}
    \centering
    \includegraphics[width=15cm]{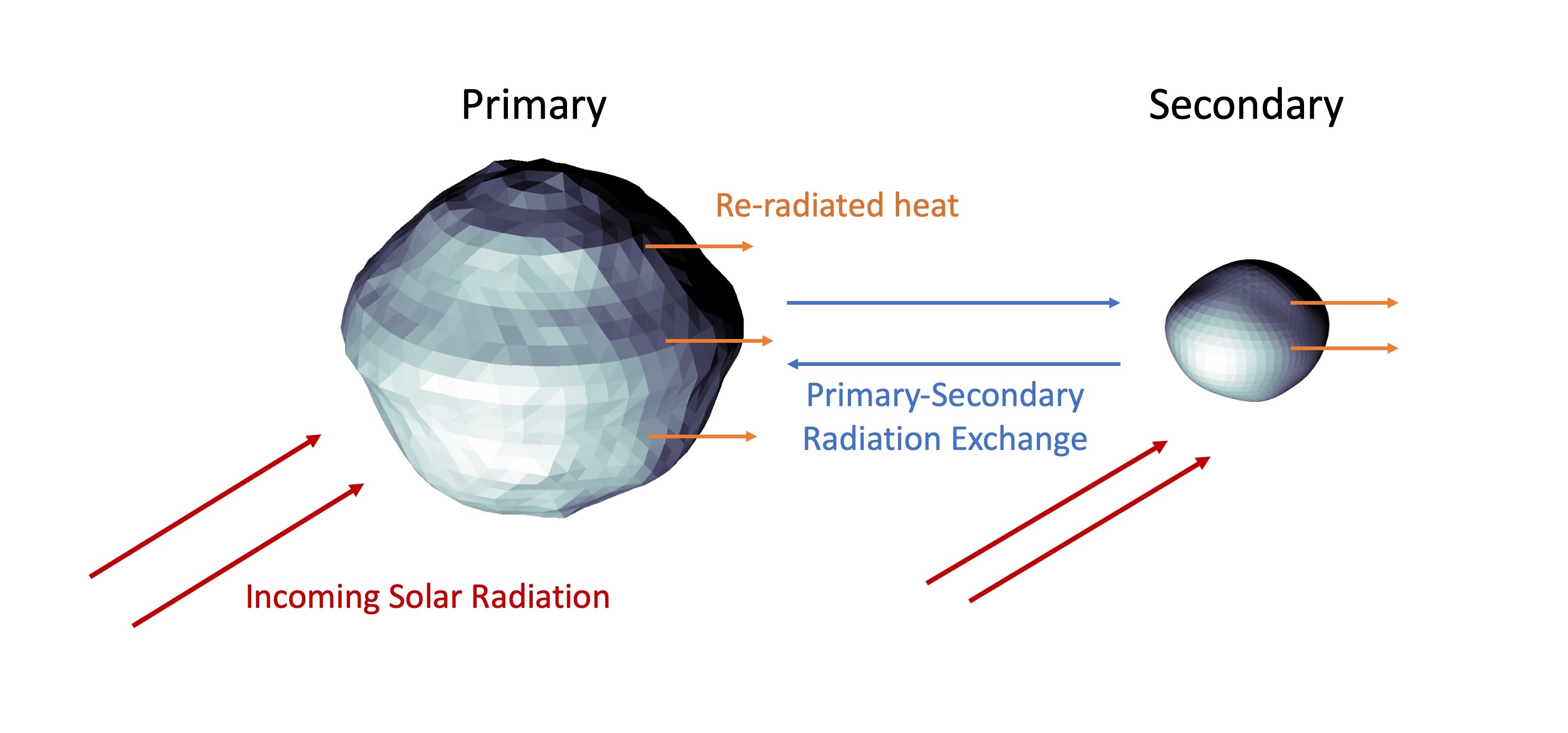}
    \caption{Binary thermophysical schematic, demonstrating flux exchange between both bodies, incoming solar radiation and re-radiated heat. Light can also be scattered between parts on the same body due to local topography.}
    \label{fig:binaryschematic}
\end{figure}

\subsection{Binary System (175706) 1996 FG3}
1996 FG3 is a binary NEA that was one of the planned targets of NASA's SIMPLEx Janus Mission \citep{Scheeres2020Janus:Asteroids, Pravec2000Two-PeriodAsteroids}. Prior to Janus, 1996 FG3 was also the suggested target for the ESA proposed MarcoPolo-R sample return mission concept. As a result, it is comparably better studied than many other binaries, with information on shapes and orbits for both bodies. It is believed to be a binary in a stable end state, with a synchronously rotating secondary \citep{Scheeres2020Janus:Asteroids}. In such a system, tidal forces are in equilibrium with the BYORP effect. This tidal-BYORP equilibrium implies that the binary mutual orbit is not secularly evolving \citep{Jacobson2011LONG-TERMASTEROIDS}. This allows the magnitude of BYORP to be estimated. Thus, 1996 FG3 is an excellent candidate for investigating binary dynamics, especially BYORP. Previous work by \cite{Scheirich2015TheEvolution} provided a measurement of the mutual orbit semimajor axis drift rate for the system at $-0.07 \pm 0.34$ cm/yr, although it is important to note that this result does not distinguish between BYORP and tidal forces. Relevant compositional and orbital parameters for 1996 FG3 are given in Table \ref{tab:fg3param}.

\subsection{Thermophysical Modeling Background}
Binary asteroid systems have complex thermophysical environments, as shown in the schematic in Figure \ref{fig:binaryschematic}. Determining which thermophysical and radiative parameters affect the magnitude of BYORP and to what extent requires a 3D thermophysical model capable of considering the radiative and dynamical interactions of the asteroid pair. 

Previous thermophysical models developed for airless bodies and solitary asteroids have often used idealized spherical geometry, thus requiring no shadowing, and simplified thermal effects like no heat conduction \citep{Lebofsky1986APallas, Harris1998AAsteroids}. However, models for airless bodies like those described in \cite{Linsky1966ModelsProperties}, \cite{Jones1975TemperaturesLayer} and \cite{Spencer1990APlanets} utilized a detailed conduction method. Later models incorporated high resolution 3D shape models to model surface temperatures, such as \cite{Groussin2007Surface1} for comet 9P/Tempel 1. Models like those produced by \cite{Delbo2004TheEmission}, \cite{Lagerros1998}, and \cite{Muller2007} used topography and realistic thermophysical effects like shadowing, 1D heat conduction and radiation exchange between facets in craters. 

Current state-of-the-art single asteroid models, such as the Advanced Thermophysical Model (ATPM) by \cite{Rozitis2011DirectionalModel}, use detailed shape models and small scale surface roughness to incorporate the illumination and emission angles for each facet. \cite{Nakano2023FiniteComputation} and \cite{Davidsson2014SurfaceModels} developed models using 3D heat conduction between facets. However, 3D heat conduction is computationally expensive and can be neglected when the size scale of the modeled topography is larger than the thermal skin depth; this is generally true for asteroids where $z_s = \sqrt{\kappa P/\pi} < 1$ cm for diffusivity $\kappa \sim 10^{-8}~\mathrm{m^2~s^{-1}}$ and period $P < 10$ hr. Thus for the relatively low resolution of shape models available for binary asteroids, 1D heat conduction is sufficient.

Previous investigations of binary asteroids with thermophysical modeling are limited. Thermal modeling for Didymos, the DART target binary, utilized rudimentary shape models and placed upper limits on the effect of primary-secondary mutual heating \citep{Pelivan2017ThermophysicalMission}. However, these Didymos models did not explicitly investigate surface topography and small-scale roughness. Recent work by \cite{Jackson2024ThermophysicalSystems} used thermal modeling, again of the Didymos system, to analyze the effect of thermal inertia on binary thermophysical light curve morphology. \cite{Yu2014ShapeFG3} performed thermophysical modeling of Janus target 1996 FG3, including consideration of thermal inertia, but assumes the subsurface beneath the regolith is a constant temperature, which can greatly limit the accuracy of diurnal surface temperature curves. 

In this paper, we describe a 3D thermophysical model we developed for binary asteroid systems that has additional broad applicability to single bodies and landscapes. This model applies and builds upon one of the most highly-validated existing planetary thermal models \citep{Hayne2017GlobalExperiment}, and adds new capabilities including ray tracing, view factor calculation, visible light reflection and IR scattering, mutual radiation and eclipsing between the binary pair. We apply the model to potential mission target system 1996 FG3 and present the following results: (1) Surface temperatures and thermal maps of the 1996 FG3 system, (2) Estimates of the magnitude of radiation exchange between the primary and secondary, and (3) Contributions of moonshine, thermal inertia and eclipses on the BYORP coefficient for the 1996 FG3 system.

\begin{table}
    \centering
    \begin{tabular}{|cccc|}
    \hline 
    \textbf{Property} & \textbf{Value} & \textbf{Unit (If Applicable)} & \textbf{Source} \\ 
    \hline
    \hline 
    \textit{System Orbital Properties} & & &\\ [0.5ex]
    \hline 
    \hline 
        Semimajor axis &  1.054 & AU & JPL Small Bodies Database\\
         \hline 
        Eccentricity &  0.3497 &  & JPL Small Bodies Database \\
         \hline 
        Perihelion &  0.6853 & AU & JPL Small Bodies Database\\
         \hline 
        Aphelion &  1.423 & AU & JPL Small Bodies Database\\
         \hline 
    \textit{Mutual Orbit Properties} & & &\\ [0.5ex]
    \hline 
    \hline 
         Semimajor axis & 2.46 & km &  \cite{Scheirich2015TheEvolution}\\
         \hline 
         Orbital Period &  16.1508 & hours & \cite{Scheirich2015TheEvolution}\\
         \hline 
    \textit{Primary Properties} & & &\\ [0.5ex]
        \hline 
        \hline 
        Mean Diameter &  1.69 $\pm$ 0.22 & km & \cite{Scheirich2015TheEvolution} \\
        \hline 
         Rotation Period &  3.595 & hours & \cite{Scheirich2015TheEvolution}\\
         \hline 
        Bulk Density &  $1400^{+1500}_{-600}$  & $\mathrm{kg / m^{3}}$ & \cite{Scheirich2009ModelingAsteroids} \\
         \hline 
    \textit{Secondary Properties} & & & \\ [0.5ex]
        \hline 
        \hline 
        Mean Diameter &  0.49 $\pm$ 0.08 & km & \cite{Scheirich2015TheEvolution}\\
        \hline 
        Rotation Period &  16.1508 $\pm$ 0.01 & hours & \cite{Scheirich2015TheEvolution}\\
        \hline
        Bulk Density &  $1400^{+1500}_{-600}$  & $\mathrm{kg / m^{3}}$ & \cite{Scheirich2009ModelingAsteroids} \\
         \hline  
    \textit{Photometric and Thermal Properties} & & & \\ [0.5ex]
        \hline 
        \hline
        SMASSII Spectral Type & C Type & & \cite{Bus2002PhaseTaxonomy} \\
        \hline
        Geometric Albedo &  0.044 $\pm$ 0.004 &  & \cite{Wolters2011PhysicalFG3} \\
         \hline 
        Bond Albedo & 0.011 && Adapted from \cite{Wolters2011PhysicalFG3} \\
        \hline 
        Thermal Inertia &  120 $\pm$ 50 & $\mathrm{J~m^{-2}~K^{-1}~s^{-1/2}}$ & \cite{Wolters2011PhysicalFG3} \\
        \hline 
        Thermal Inertia &  80 $\pm$ 40 & $\mathrm{J~m^{-2}~K^{-1}~s^{-1/2}}$ & \cite{Yu2014ShapeFG3} \\
        \hline 
        Ryugu Thermal Inertia &  225 $\pm$ 45 & $\mathrm{J~m^{-2}~K^{-1}~s^{-1/2}}$ & \cite{Shimaki2020ThermophysicalMapping} \\
        \hline 
        Emissivity & 0.95 & & Assummed \\
        \hline
    \end{tabular}
    \caption{Table of pertinent parameters for binary system 1996 FG3. For parameters where more than one value has been published, such as thermal inertia, both are given. Pole positions are given in ecliptic coordinates.}
    \label{tab:fg3param}
\end{table}


\section{Methods} \label{sec:methods}

\subsection{Binary Thermophysical Model}

In this section, we introduce the Binary Thermophysical Model (BTM), which calculates temperatures for a 3D binary system, but can also be applied to solitary asteroids and surfaces. The BTM considers subsurface heat conduction and uses 3D surfaces and ray tracing to include topography, shadowing, infrared emission and scattering, as well as visible light reflection. 

\subsubsection{The 1D Model}
To incorporate subsurface heat transfer, we begin with a 1D thermophysical model from \cite{Hayne2017GlobalExperiment}. The model solves the 1D heat diffusion equation:

\begin{equation}
    \rho c_{\mathrm{p}}\frac{\partial T}{\partial t} = \frac{\partial}{\partial z}\left(k\frac{\partial T}{\partial z}\right)
\end{equation}
where $T = T(z)$ is the temperature, $k = k(z, T)$ is the thermal conductivity ($\mathrm{W~m^{-1}~K^{-1}}$)), $c_{\mathrm{p}} = c_{\mathrm{p}}(T)$ is the specific heat capacity ($\mathrm{J~kg^{-1}~K^{-1}}$) and $\rho = \rho(z)$ is the density of the medium ($\mathrm{kg~m^{-3}}$). The model is capable of specifying distinct thermophysical parameter values, including density and thermal conductivity, at each spatial grid point.

 The numerical thermal model applies a standard finite difference approximation for the spatial and temporal derivatives \citep{Hayne2015ThermalTopography}. At the surface, the model balances incident radiation from insolation and scattered radiation, with conduction and infrared emission. This boundary condition is solved using Newton’s root finding method at each time step. Heat flow at the lower boundary is set to zero. For further details on the numerical model, see Appendix A of \cite{Hayne2017GlobalExperiment}. For specific thermophysical properties used by the model, see Table \ref{tab:fg3param}.

Building from this particular model has several advantages: (1) we account for temperature-dependent thermal conductivity and specific heat capacity, which can vary by orders of magnitude over typical diurnal cycles near 1 AU \citep{Hayne2015EvidenceMeasurements};  (2) we account for depth-dependent density/porosity profiles, which can have substantial effects on diurnal surface temperature curves (e.g., \cite{Vasavada2012LunarExperiment}); (3) this 1D model has been rigorously validated for airless bodies against the global thermal IR Lunar dataset from the Diviner Lunar Radiometer experiment (e.g., \cite{Williams2019SeasonalMoon}). Additionally, modifications to address the effects of sub-pixel-scale roughness, e.g. "IR beaming", can be incorporated as required \citep{Bandfield2015LunarObservations, Rozitis2011DirectionalModel}. 

Various applications of the model, including interpretation of in-situ thermal data, observation planning, and prediction of the magnitude of the BYORP effect, are sensitive to small changes in temperature. Thermal inertia can induce significant changes in the range and duration of temperatures experienced on airless bodies. Accordingly, a focus of this work is to consider a range of thermal inertia values on the target system 1996 FG3, which can eventually be used to constrain the real system values. Thermal inertia, $\Gamma$, depends on the thermal conductivity $k$,  bulk density $\rho$, and the specific heat capacity $c_{\mathrm{p}}$. 

\begin{equation}
\Gamma = \sqrt{k \rho c_{\mathrm{p}}}
    \label{eq:inertia}
\end{equation}

The BTM considers variations of conductivity with depth and temperature as well as the dependence of heat capacity on temperature \citep{Ledlow1992SubsurfaceCentimeters}, although the latter is expected to be a smaller effect than thermal conductivity for asteroids \citep{Rozitis2020AsteroidEquator}. The temperature dependence of the thermal conductivity takes the form $k(T) = k_\mathrm{c} + \beta T^3$, where $k_\mathrm{c}$ is the temperature-independent conductivity relevant at low temperatures, and $\beta$ is a constant of order $10^{-11}~\mathrm{W~m^{-1}~K^{-4}}$ \citep{Whipple1950}. The temperature-dependent term in the thermal conductivity dominates for temperatures higher than $\sim 250~\mathrm{K}$. Thus, the BTM is capable of representing how density varies with depth, specific heat varies with temperature and thus with time, and conductivity varies with both depth and time. As a result, thermal inertia is a depth- and time- dependent property in our formulation \citep{Hayne2017GlobalExperiment}, although this can be turned off and thermal inertia considered constant with depth and temperature as needed.

The strength of the temperature and depth dependence of $\Gamma$ on asteroids is not well known \citep{Rozitis2020AsteroidEquator}. These effects should be important for fine grained regolith such as that found on the Moon \citep{Hayne2017GlobalExperiment}, but less important for asteroids with higher $\Gamma$ values (hence lower porosity). However, as current thermal inertia estimates for 1996 FG3 are considerably lower than the values found for asteroids like Bennu and Ryugu (see section \ref{FG3ParamSection}), we include the temperature dependence here. To simplify the presentation of our results, in what follows we report the thermal inertia of the surface layer neglecting the temperature-dependent terms; otherwise, a range of values would need to be given representing the full diurnal cycle and latitude dependence.

Variations of thermal inertia and specific heat with depth are also likely to be small, as density variations with depth are expected to be minimal for asteroid surfaces \citep{Rozitis2020AsteroidEquator, Bierhaus2023AEvolution}. The thermal skin depth $z_{\mathrm{S}} = (k P / \pi \rho c_{\mathrm{p}})^{1/2}$, where $P$ is the rotation period, is the depth to which the diurnal heat wave propagates. For 1996 FG3, this is expected to be shallow, likely around 1-2 cm for the primary and 3-4 cm for the slower-orbiting secondary. Thus, again the thermal inertia values reported in what follows are the values for the surface layers.

\subsubsection{3D Ray Tracing}
Topography can significantly alter temperatures on airless bodies. In the complex thermophysical environments of binary systems, this includes self-shadowing, radiation exchange between the primary and secondary, and the shadowing induced by eclipses. We incorporate these effects through a custom ray tracer as well as observational and shape data.

To represent these 3D effects, we couple the 1D model to asteroid shape models derived from from Earth-based radar data (Arecibo). These shape models are supplied by the Janus mission team (L. A. M. Benner, personal communication), and are available through the link provided to the BTM. By integrating the model with Trimesh \citep{Dawson-Haggertyetal.2019}, the BTM is capable of loading one or more triangular meshes, such as .obj, .wf or .stl files. This allows the BTM to be applied to surfaces or landforms, solitary small bodies or multiple systems. 

Inputs to the model include system orbital parameters such as semimajor axis, eccentricity, and obliquity to enable calculation of the heliocentric distance and the solar constant at each timestep \citep{Hayne2015ThermalTopography}. The BTM also allows for control over binary orbital parameters, including primary rotation period, secondary mutual orbit period and spatial separation between the primary and secondary. 

By tracing rays from the Sun to each facet at each time step, we determine which facets are in shadow or eclipsed. The flux incident on a facet ($Q_{\mathrm{i}}$) is dependent on whether the object is in shadow or not ($V_{\mathrm{i}} = 0$ if shadowed, 1 if illuminated), the distance from the sun $R$, the solar vector $\hat{S_{\mathrm{i}}}$ from the Sun to the facet and its dot product with the normal vector for the facet $\hat{n_{\mathrm{i}}}$. The insolation on the $i^\mathrm{th}$ facet can thus be represented by Eq. \ref{eq:insolation}, where $L_{\odot}$ is the solar luminosity and $\frac{L_{\odot}}{4\pi R^{2}}$ is the irradiance at a distance $R$, in W/$m^{2}$.

\begin{equation}
    Q_{\mathrm{i}} = V_{\mathrm{i}} \frac{L_{\odot}}{4\pi R^{2}} \hat{S_{\mathrm{i}}} \cdot \hat{n}
    \label{eq:insolation}
\end{equation}

Ray tracing also allows for precise calculation of inter-facet reflection and emission through the so-called view factors. The view factor $F_{\mathrm{ij}}$ is defined as the fraction of radiation from a facet, $i$, that reaches another facet, $j$ \citep{Rezac2020AstronomyAsteroids}.  For discrete areas $A_{\mathrm{i}}$ and $A_{\mathrm{j}}$,  the simple form of this is shown in Equation \ref{eq:simpleVF}. This relates the view factor with the cosine of the angles, $\phi_{\mathrm{ij}}$ and $\phi_{\mathrm{ji}}$, between the normal of the each facet with the vector between the facets. 

\begin{equation}
    F_{\mathrm{ij}} = \frac{A_{\mathrm{j}} \cos\phi_{\mathrm{ij}} \cos\phi_{\mathrm{ji}}}{\pi d^{2}_{\mathrm{ji}}}
    \label{eq:simpleVF}
\end{equation}

We implement the "M2" method from \cite{Rezac2020AstronomyAsteroids}, which is more robust if the distance $d_{\mathrm{ij}}$ between facets is small compared to the area of the facets. This method is reproduced in Equation \ref{eq:m2VF}.

\begin{equation}
    F_{\mathrm{ij}} = \frac{4\sqrt{A_{\mathrm{i}} A_{\mathrm{j}} }}{\pi^{2} A_{\mathrm{i}}}\arctan \left[\frac{\sqrt{\pi A_{\mathrm{i}}}\cos\phi_{\mathrm{ij}}}{2 d_{\mathrm{ij}}}\right] \arctan \left[\frac{\sqrt{\pi A_{\mathrm{j}}}\cos\phi_{\mathrm{ji}}}{2 d_{\mathrm{ij}}}\right]
    \label{eq:m2VF}
\end{equation}

We use single scattering for all view factor calculations. Multiple scattering does not cause a significant effect on temperatures given the low albedo of 1996 FG3. After two reflections, the amount of radiation scattered is proportional to the albedo squared. Given a Bond Albedo of 0.011 (See Table \ref{tab:fg3param}), less than 0.1\% of the incident insolation is reflected after a second scatter event.     

The view factors between facets on a single shape model need only be calculated once. However, the dynamic system geometry of binary systems requires updating shadows and view factors between independently moving bodies, which can lead to high computational costs. We implemented several techniques to reduce the computational costs of dynamic ray-tracing. First, the model uses multiprocessing to run many models in parallel. Second, the BTM allows for usage of pre-calculated and adaptive ray tracing. For tidally locked systems like 1996 FG3, the same face of the secondary always faces the primary. We pre-calculate shadows and view factors for every potential location of the secondary about the primary, and create a look up table that can be referenced by the BTM to find the closest orientation at each timestep of the thermal model. To ensure the look up table is sufficiently detailed for interpolation, we iteratively increased the resolution of the pre-calculated tables until further increases no longer produced significant differences in the results.

Though this approach requires many initial computations, the BTM can then use these results for all future runs using the same binary orbital parameters, greatly reducing long-term computational cost. This is especially true when many time steps are needed to ensure model stability. The BTM can also perform adaptive ray tracing at each time step, or at a given interval, to give exact results for the current geometry. This is applied after equilibration and during BYORP calculation, and will eventually be used for dynamically evolving systems.

\subsection{Model Validation}

As in-situ temperature data have yet to be obtained for a binary system, we validated the BTM with analytical solutions and spacecraft data for single bodies. To validate view factor and shadowing calculation, we compared BTM results to exact solutions for instantaneous surface temperatures of an equatorial bowl-shaped crater on the Moon from \cite{Ingersoll1992StabilityMars}. For instantaneous temperatures, our ray-tracing model matches predicted temperatures in a bowl shaped crater to better than 1 K.

\begin{figure}
    \centering
    \includegraphics[width=18cm]{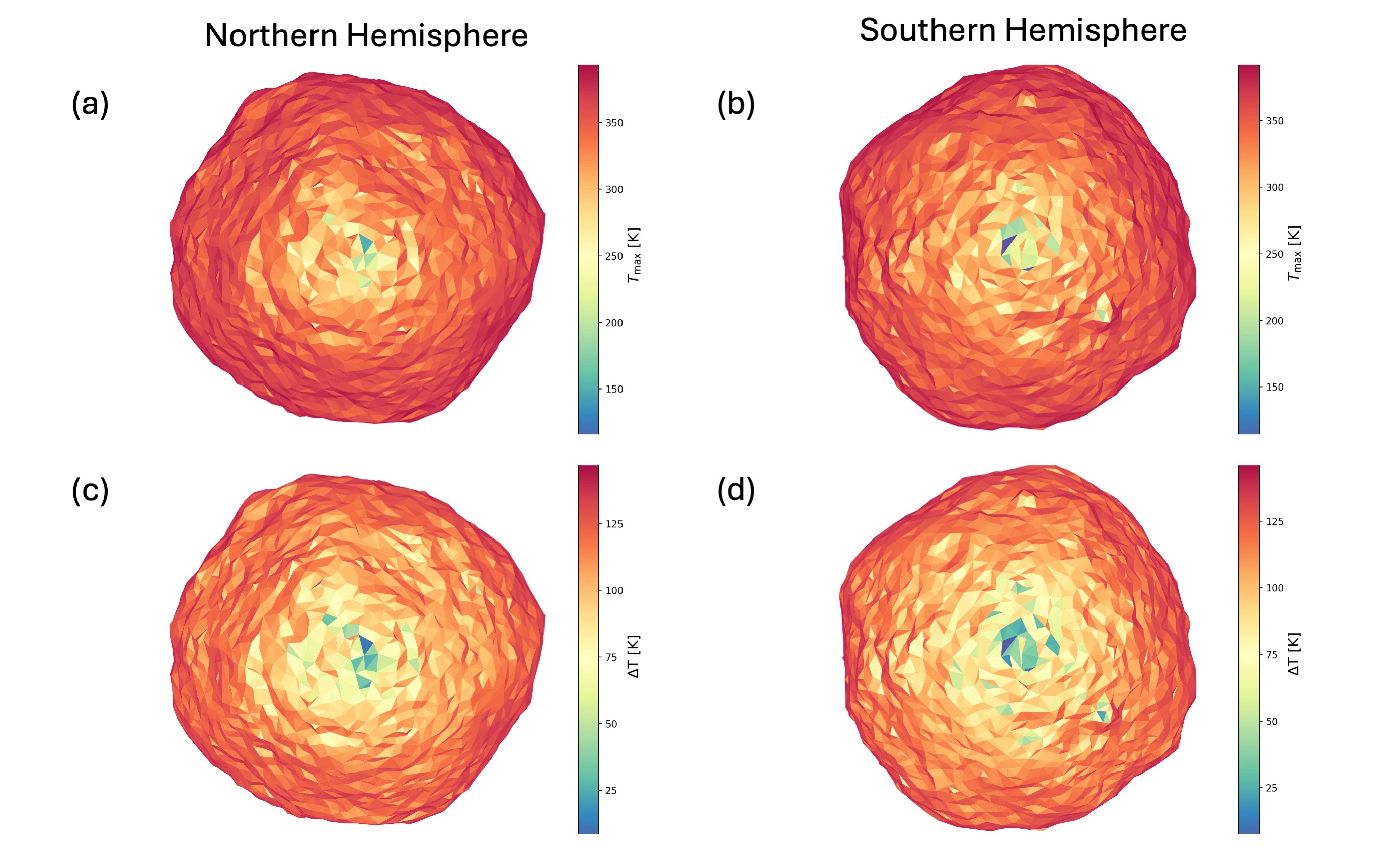}
    \caption{Maximum facet temperatures (panels a, b), and diurnal amplitudes (panels c, d) for the northern and southern hemispheres of Bennu using the BTM.}
    \label{fig:BennuVal}
\end{figure}

We also compared the BTM against observational data and existing thermophysical modeling of OSIRIS-REx target asteroid (101955) Bennu from \cite{Rozitis2020ImplicationsBennu} and \cite{Rozitis2020AsteroidEquator}, similar to the approach utilized by \cite{Nakano2023FiniteComputation}. We begin with the 75-cm resolution Global Digital Terrain Map (GDTM v20) generated from data from the OSIRIS-REx spacecraft \citep{Barnouin2020DigitalMission}. As most binary shape models are only a few thousand facets at most, we reduce the resolution of this shape model from roughly 3 million facets to about 5900.  We assume uniform thermophysical properties for Bennu, including a thermal inertia of $~350$ tiu (thermal inertia units, or $\mathrm{J~m^{-2}~K^{-1}~s^{-1/2}}$), and orbital parameters given by Table 2 in \cite{Nakano2023FiniteComputation}. To compare with the model described by \cite{Rozitis2020AsteroidEquator} and the validation results presented in \cite{Nakano2023FiniteComputation}, we also turned off thermally dependent conductivity and assumed no secondary scattering. Using these parameters, we find a global maximum temperature of 399 K, an average maximum temperature of 356 K, an average diurnal amplitude of 113 K and a minimum temperature of 102 K. These values are consistent with those shown in Figure 3 of \cite{Rozitis2020ImplicationsBennu}, and the global maximum T is within 1 K of that derived by \cite{Nakano2023FiniteComputation}. Note that each validation effort used a different artificially simplified shape model of Bennu, and ours has roughly 18\% more facets than that used by \cite{Nakano2023FiniteComputation}. Maps of our maximum temperatures and diurnal amplitudes are shown in Figure \ref{fig:BennuVal}.

Finally, we performed an energy balance analysis of the full binary model, which shows that no significant energy leaks exist. 

\subsection{BYORP Calculation} \label{sec:BYORP}
Here, we use the BTM to calculate the magnitude of the BYORP effect, and investigate how binary effects alter the BYORP coefficient $B$. $B$ can vary between 0 for a symmetric body and a maximum value of 2, with most binary systems having a $B \sim 10^{-3}$ \citep{Jacobson2011LONG-TERMASTEROIDS}. 

Using the standard BYORP model from \cite{McMahon2010SecularBYORP}, based on the methodology from \cite{Scheeres2007TheYORP}, we calculate a theoretical value of $B_{o} = -0.0186976$ for the given shape models. This model does not account for eclipses, thermal inertia or mutually exchanged radiation. We compare this theoretical value against $B$ values produced by the BTM with those properties. 

The BTM calculates fluxes and temperatures that can be used to quantify the instantaneous radiation pressures and forces from solar radiation and re-emission. Previous work indicates there is no secular change to BYORP as a result of absorption of radiation \citep{Rubincam2010}. Existing BYORP theory thus includes the effects of reflection and emission, but not absorption. At present, our model does not include absorption but this component could be added. The force acting on the $i^\mathrm{th}$ facet can be described by Equation \ref{eq:radpressure}. The $1/c$ term comes from radiation pressure $P = \Phi / c $. Here $P$ is the magnitude of the pressure (Pa), $\Phi$ is the irradiance incident on a surface ($\mathrm{W~m^{-2}}$), and $c$ is the speed of light ($\mathrm{m~s^{-1}}$). For a perfectly absorbing surface perpendicular to the incoming solar vector, $\Phi$ is equal to the base solar insolation at a given heliocentric distance. However, $\Phi$ is dependent on several factors in the BTM, including albedo, visibility, shape and in the case of self heating due to re-emission, temperature of neighboring facets. 

To more accurately approximate asteroid surfaces, reflected visible light and thermal emission are modeled as diffuse. Radiation is reflected in all directions, consistent with Lambert's Law of Diffusion. In this regime, tangential components cancel and the remnant force is directed anti parallel to the surface normal, or $-\hat{n_{\mathrm{i}}}$ direction. The associated Lambertian reflectance factor is 2/3. $A_{\mathrm{i}}$ is the area of the facet. The first term describes the recoil force from reflecting flux, given by the bond albedo times the intensity incident on the facet, or $Q_{\mathrm{ref}} = A_{B} Q_{\mathrm{insolation}}$. This includes both insolation and scattered light from other facets. The second term describes the force arising from thermal emission, where $\sigma$ is the Stefan Boltzmann Constant and $\epsilon$ is the emissivity. 

\begin{equation}
F_{\mathrm{i}} = -\frac{2}{3c} A_{\mathrm{i}} \hat{n_{\mathrm{i}}} \left(Q_{\mathrm{ref}} + \sigma \epsilon T_{\mathrm{i}}^{4}\right)
\label{eq:radpressure}
\end{equation}

We sum these force vectors over the whole body. To compare with the theoretical value of $B_{\mathrm{o}}$ \citep{Scheeres2007TheYORP, McMahon2010SecularBYORP}, equal to the $A_{0(2)}$ coefficient, we are concerned with the force in the y-direction in the secondary-fixed frame, or the y component of the facet force vectors. We solve for $B$ using Eq. \ref{eq:a02}. For each model timestep, we remove the heliocentric distance dependence by dividing by P(R), the solar pressure at the exact distance of the system from the Sun at that timestep. We integrate over all timesteps in a mutual orbit. This is represented in Eq. \ref{eq:a02} as an integral over the entire mutual orbit with respect to the longitude of the sun in the secondary's body frame, $\lambda_{\mathrm{s}}$. The unitless, normalized $B$ is obtained by dividing by the square of the volume equivalent sphere radius ($R_{\mathrm{RES}}^{2}$) times $2\pi$.

\begin{equation}
    B \equiv A_{0(2)} = \frac{1}{2\pi R_{\mathrm{RES}}^{2}}\int_{0}^{2\pi} \frac{F_{\mathrm{y}}(t)}{P(R)} d\lambda_{\mathrm{s}}
    \label{eq:a02}
\end{equation}.

For each BTM run, we calculate the BYORP Coefficient $B$ and compare it to the theoretical value.


\subsection{1996 FG3 Parameters \& Model Runs} \label{FG3ParamSection}
We apply the BTM to the stable, singly-synchronous binary, 1996 FG3. The shape models for the 1996 FG3 primary and secondary are each composed of 2292 facets. Additional inputs to the BTM for these runs are given in Table \ref{tab:fg3param}. Knowledge of the binary mutual orbit is limited. In addition to the values shown in Table \ref{tab:fg3param}, \cite{Scheirich2015TheEvolution} found a maximum possible eccentricity for the mutual orbit of $e = 0.07$. As this upper limit is low, and the actual value even lower, we assume a circular mutual orbit for simplicity. We also assume zero inclination of the mutual orbit as this value is unconstrained \citep{Scheirich2015TheEvolution}.

Using a geometric albedo of 0.044 from \cite{Wolters2011PhysicalFG3}, we estimate a bond albedo value of 0.011 using a phase integral value of 0.25. Due to the small size of the bodies, we neglect geothermal heat flux. \cite{Wolters2011PhysicalFG3} and \cite{Yu2014ShapeFG3} have estimated the thermal inertia to be $\Gamma = 120 \pm 70$ tiu and $\Gamma = 80 \pm 40$ tiu, respectively. These low values indicate a dusty, highly fragmented regolith layer. Both thermal inertia values are smaller than the in-situ value obtained for other C-type asteroid Ryugu of $\Gamma = 225 \pm 45$ tiu \citep{Shimaki2020ThermophysicalMapping}.  

Thermal inertia of secondaries in a binary pair is a largely unconstrained parameter because primaries dominate observations. Several studies, such as \cite{Delbo2009ThermalData}, have analyzed the dependence of thermal inertia on asteroid size, with smaller asteroids expected to have higher thermal inertias. As many secondaries are believed to form via spin up of the original parent body, they may have thermophysical properties similar to their primaries. However, the extremely low gravity and presence of electrostatic charge on small secondaries may contribute to particle lofting, potentially transferring material between binary components and either stripping or covering the secondary with fine to medium grained material \citep{Maruskin2013DustSystems}. We thus consider a range of possible thermal inertias: 0, 40, 80, 160 and 320 tiu. 

In addition, to further investigate the sensitivity of BYORP associated with thermal inertia, we also run the BTM with the high thermal inertia values of 700 tiu (Itokawa-like from \cite{Muller201425143Results}) and 2500 tiu (bare rock from \cite{Jakosky1986OnFines}). Though these values are not realistic for 1996 FG3, they are useful in demonstrating the trends in $B$ associated with increases in thermal inertia.

We performed over 40 BTM runs for this study. These runs were clustered into the following groups: 
\begin{itemize}
    \item \textbf{Group 1: Theoretical Comparison}: Zero thermal inertia for both bodies, with no eclipses and no mutual radiation.
    \item \textbf{Group 2: Secondary Thermal Inertia Effect}: Holding the primary thermal inertia at 160 tiu, we run the secondary at the mentioned range of thermal inertia, with eclipses and mutual radiation off. 
    \item \textbf{Group 3: Eclipse Effect}: Same as Group 2, but with eclipses enabled 
    \item \textbf{Group 4: Mutual Radiation Effect}: Same as Group 2, but with eclipses and mutual radiation enabled 
    \item \textbf{Group 5: Primary Thermal Inertia Effect on Mutual Radiation}: Secondary thermal inertia held at 160 tiu and the primary is modeled at a range of thermal inertias, with mutual radiation and eclipses enabled. This allows consideration of how primary properties alter the strength of mutual radiation from the primary to the secondary.
\end{itemize}

Each BTM run returns surface and subsurface temperatures, as well as the BYORP coefficient $B$, measured at perihelion. Due to the 1996 FG3 system's high eccentricity ($e > 0.3$), temperatures are higher at perihelion, and insolation decreases by about $75\%$ at aphelion. The zeroth order Fourier coefficients used to quantify BYORP are considered independent of location along heliocentric orbit. However, if the magnitude of BYORP is sensitive to thermophysical properties, we expect it to be most pronounced near perihelion.

\section{Results}
Here we present the results produced by the BTM and BYORP model, including temperature metrics, surface temperature maps, the effect of mutual radiation between the binary pair and BYORP coefficients. 

\subsection{Binary Thermal Model Results}
Temperature results from BTM model runs for the 1996 FG3 system at perihelion are summarized in Table \ref{tab:metrics}. These results are shown as global temperature metrics, including the maximum temperature, minimum temperature, mean temperature and diurnal amplitude for each body. See section \ref{sec:moonshineresults} for the results on temperature of mutual radiation. Note that all reported $\Gamma$ values are for the thermal inertia at the surface. As expected, increased thermal inertia reduces the diurnal temperature amplitude experienced by each facet, as well as the maximum temperature. It also generally leads to higher minimum temperature and mean temperature, and does so consistently for the range of thermal inertia values represented in the general small body population (ie. $\Gamma = 0 - 700$ tiu). However, for the rock-like thermal inertia of  $\Gamma = 2500$ tiu, there is a levelling off of the mean temperature and slight decrease in minimum temperatures compared to the $\Gamma = 700$ tiu case. The strong resistance to heating and short asteroid rotational periods reduce the number of warm facets that can provide heating from scattering to shadowed polar facets.

    \begin{table}
    \begin{center}
    \begin{tabular*}{0.69\textwidth}{||c || c  c  c c || c c||}
    \hline 
    \rule{0pt}{4ex}    \textbf{ Thermal Inertia (tiu)} & $T_{\mathrm{max}}$ & $T_{\mathrm{min}}$ & Mean T & $\Delta T$ & $M_{\mathrm{Mean}}$ & $M_{\mathrm{Max}}$ \\
     \hline
     \hline
    \rule{0pt}{4ex} \textbf{1996 FG3 Primary} &  &   &  & & & \\ 
     \hline 
     0 & 477 K & 86 K & 277 K &  291 K &  -- & -- \\ 
     \hline
    40 & 471 K & 100 K & 295 K &  252 K &  0.034 \% & 0.16 \%\\ 
     \hline
    80 & 462 K & 110 K & 307 K & 211 K &  0.032 \% & 0.14 \&\\
     \hline
    160 & 447 K & 117 K & 319 K & 164 K & 0.030 \% & 0.12 \% \\
     \hline
    320 & 425 K & 120 K & 327 K & 114 K & 0.027 \% & 0.11 \% \\
    \hline
    700 & 401 K & 120 K & 331 K & 65 K & 0.023 \% & 0.083 \%\\
    \hline
    2500 & 373 K & 111 K & 330 K & 21 K & 0.018 \% &  0.082 \%\\
    \hline 
    \hline
    \rule{0pt}{4ex} \textbf{1996 FG3 Secondary} &  &  & & & &\\
     \hline
     0 & 478 K & 85 K & 249 K &  324 K & -- & -- \\ 
     \hline
    40 & 475 K & 101 K & 265 K &  292 K & 1.35 \% & 14.3 \%\\ 
     \hline
    80 & 470 K & 115 K & 280 K & 258 K & 0.66 \% & 4.86 \%\\
     \hline
    160 & 461 K & 126 K & 294 K & 218 K & 0.38 \% & 3.0 \% \\
     \hline
    320 & 447 K & 134 K & 306 K & 170 K & 0.24 \% & 1.9 \% \\
    \hline
    700 & 422 K & 136 K & 314 K & 113 K &  0.16 \% & 1.5 \% \\
    \hline
    2500 & 380 K & 132 K & 311 K & 44 K &    0.10 \% & 0.97 \% \\
    \hline
    \hline
    \end{tabular*}
    \caption{Perihelion temperature metrics for both the 1996 FG3 primary (top) and secondary (bottom). The center section shows global temperature values, including the maximum temperature, minimum temperature, mean temperature, and diurnal amplitude for the entire body over one mutual orbit. The metrics are shown for each of the modeled thermal inertias, including 0 (the BYORP validation case), 40 - 320 (estimates for the 1996 FG3 system) as well as the high thermal inertia values (700, 2500) used to consider the effect of high thermal inertia on BYORP. The right section shows metrics related to mutual radiation between the bodies, including the mean percent change in temperature due to mutual radiation ($M_{\mathrm{Mean}}$) and the maximum percent change in temperature due to mutual radiation ($M_{\mathrm{Max}}$).} 
    \label{tab:metrics}
    \end{center}
    \end{table}

    \begin{figure}
    \centering
    \includegraphics[width=18cm]{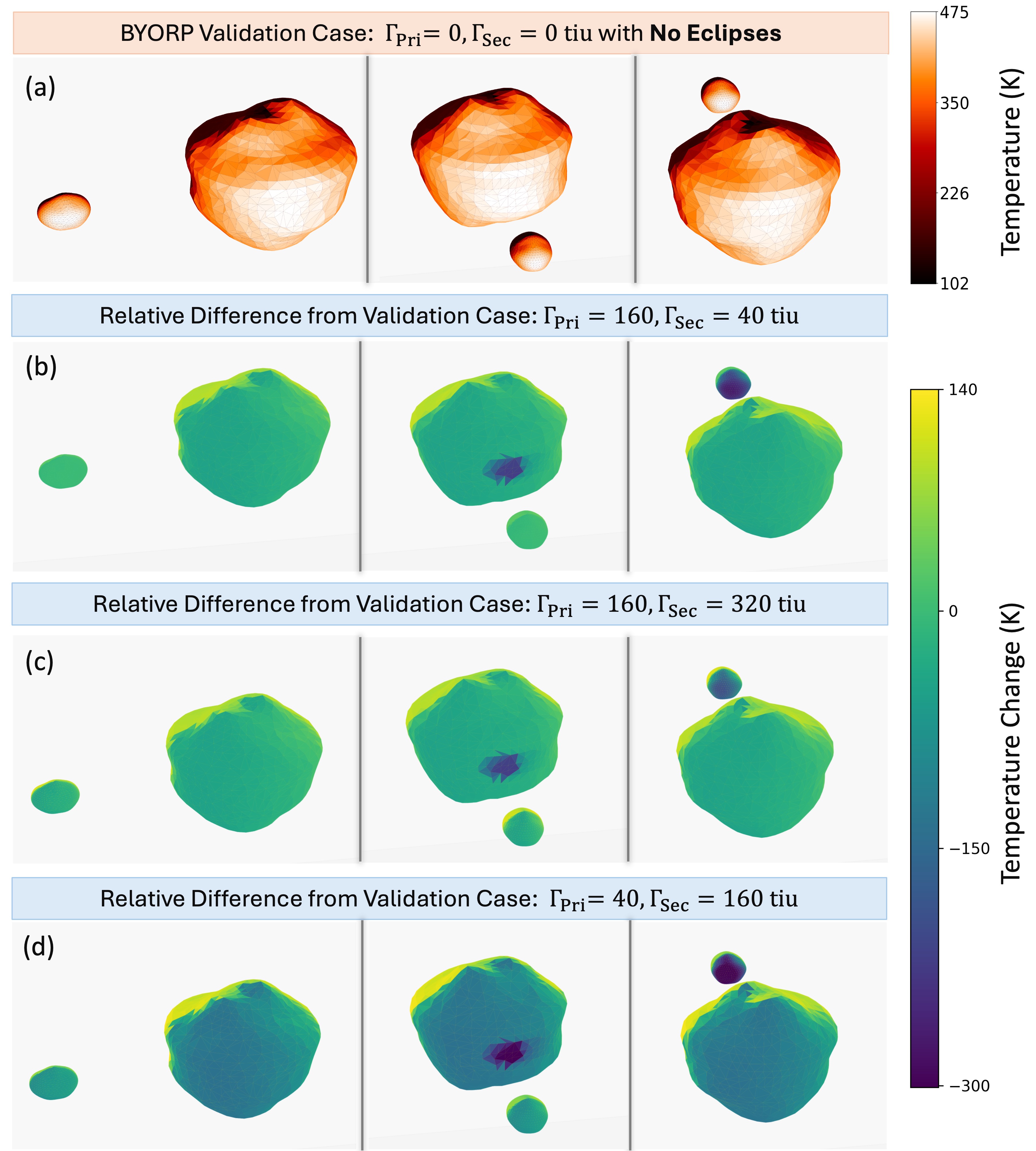}
    \caption{Surface temperature maps for a selection of model runs with varying surface thermal inertias. From left to right, each row shows the progression of the secondary about the primary, first in a state with no eclipses (leftmost column), an eclipse of the primary by the secondary (center column) and a total eclipse of the secondary (rightmost column). The top panel, labeled (a), shows modeled temperatures for the case used to compare against BYORP theory, including zero thermal inertias for both bodies, no eclipses and no mutual radiation. Panels (b) through (d) show the difference in surface temperatures relative to the validation case (panel (a)) when realistic binary effects such as eclipsing and mutual radiation are included. Panel (b) demonstrates a secondary thermal inertia of $\Gamma = 40$ tiu and a primary thermal inertia of $\Gamma = 160$ tiu. (c) maintains $\Gamma = 160$ tiu for the primary, with $\Gamma = 320$ tiu for the secondary, on the high end of the estimated range for 1996 FG3's thermal inertia. 320 tiu indicates a stronger, less fluffy material for the secondary relative to panel (b). In the final row, (d) shows a case with thermal inertia of $\Gamma = 40$ tiu for the primary, lower than the two higher panels, and a central value of $\Gamma = 160$ tiu for the secondary. Note that as the thermal inertia increases, the range of surface temperatures is dampened.}
    \label{fig:temperaturemaps}
\end{figure}

The results of several sample runs are shown in Figure \ref{fig:temperaturemaps}. These include surface temperatures for the BYORP validation case (a), which uses $\Gamma = 0$ tiu for both bodies and does not include eclipses or mutual radiation. For panels b through d, the binary effects of eclipsing and mutual radiation are included. The color indicates the difference in temperature of the model run relative to the BYORP validation case shown in (a), with darker colors indicating a cooler temperature and brighter colors a warmer temperature. 

For the middle two panels (b, c), the primary is held at a central thermal inertia of $\Gamma = 160$ tiu. The secondary's thermal inertia is shown at the low end ($\Gamma = 40$ tiu) and high value ($\Gamma = 320$ tiu) of the estimated thermal inertia range for 1996 FG3. In the final panel, the primary is shown with a low thermal inertia of $\Gamma = 40$ tiu, and the secondary has a central thermal inertia of $\Gamma = 160$ tiu. As expected, higher thermal inertia dampens the diurnal amplitude, leading to more consistent temperatures over the secondary as $\Gamma$ increases.  For runs with nonzero thermal inertia, the night sides on both bodies are warmer relative to the validation case, and the day sides are slightly cooler. Additionally, comparison of the BYORP validation case against the three runs with realistic binary effects demonstrate the marked effect eclipses have on surface temperatures. This is especially true for the totally eclipsed secondary, which has a temperature asymmetry where half the facets never reach a peak temperature comparable to that on the other side of the body. Eclipsed areas on the primary are also much cooler.  

\begin{figure}
    \centering
    \includegraphics[width=18cm]{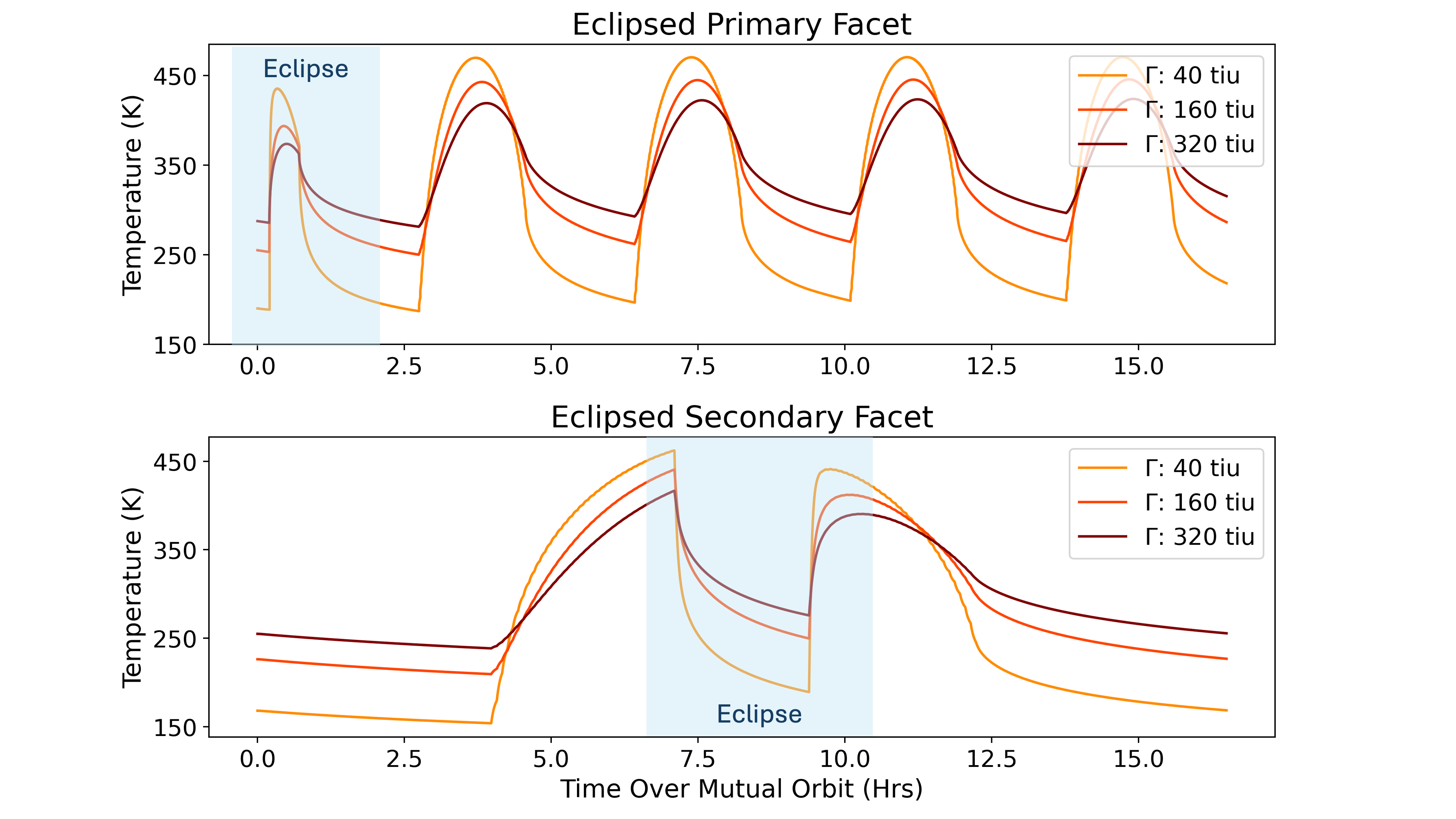}
    \caption{Diurnal temperature curves for a single facet that experiences eclipsing on (a) the primary and (b) the secondary. Each shows a thermal inertias of $\Gamma = $ 40, 160 and 320 tiu, encompassing a range of estimated values for 1996 FG3 (see Table \ref{tab:fg3param}). Note that the primary rotates multiple times during a single mutual orbit. For the selected facet, the first diurnal cycle is affected by a short eclipse by the secondary, abruptly reducing the temperature, as well as the peak temperature reached in that cycle. It also does not repeat, as the facet is not eclipsed during the following diurnal cycles.  }
    \label{fig:diurnaleclipsecurves}
\end{figure}

The synchronous rotation of the secondary leads to the same hemisphere of facets being eclipsed during every mutual orbit. In contrast, the freely rotating primary rotates a non-integer number of times during each mutual orbit. As a result, the facets eclipsed on the primary vary with each mutual orbit, and a facet that is eclipsed will not be eclipsed during the subsequent primary rotation. This causes a an irregular drop in temperature. As the facets that are eclipsed on the primary are inconsistent between rotations, the net effect on global temperature metrics is small for the primary, with a mean temperature change of less than 1 K. In contrast, the secondary experiences a larger net effect. In the most extreme case of 0 thermal inertia, eclipses reduce the mean temperature of the secondary's surface by about 20 K. Eclipses can cause a drop in temperature that ranges from a few degrees, as in the case of a facet on the periphery of an eclipse, to over 100 K for totally eclipsed equatorial facets on the secondary. For model runs with low thermal inertia values of $\Gamma = 40$ or 80 tiu, this drop can reach over 200 K. The effect of eclipses is demonstrated more explicitly in Figure \ref{fig:diurnaleclipsecurves}, which shows the diurnal curves for two eclipsed facets - one on the primary and one on the secondary - within the range of thermal inertias estimated for the 1996 FG3 system (40 - 320 tiu).

\subsection{Effect of Mutual Radiation} \label{sec:moonshineresults}
Radiation scattered between the primary and secondary can affect surface temperatures. We quantify this effect by comparing temperatures at each time step between model runs with equivalent thermophysical properties, but with mutually exchanged radiation between the primary and secondary turned on or off. A useful metric is the percent change in temperature from mutual radiation, $M$, which can be represented as the change in temperature due to mutual radiation divided by the temperature without mutual radiation ($T_{\mathrm{o}}$): 
\begin{equation}
    M = \frac{\Delta T}{T_{\mathrm{o}}} \times 100 
\end{equation}

The maximum percent change due to mutual radiation can thus be represented: 
\begin{equation}
    M_{\mathrm{max}} = \frac{\mathrm{Max}(\Delta T)}{T_{\mathrm{o}}} \times 100 
\end{equation}

The values for the $M_{\mathrm{Max}}$, as well as the mean change in temperature due to mutual radiation ($M_{\mathrm{Mean}}$) are shown on the right side of Table \ref{tab:metrics}. The secondary experiences 1-2 orders of magnitude more heating from mutual re-radiation from the primary. The maximum temperature change to the primary as a result of radiation from the secondary is $<1\%$. In contrast, the magnitude of heating on the secondary is larger and varies strongly with thermal inertia. For a thermal inertia of 40 tiu, the mean percentage change due to mutual radiation from the primary is over 1 \%, or about 3 K, and is nonzero for the facets that continually face the primary. The maximum change to a facet's temperature due to mutual radiation is 14.3\%, or about 22 K. This falls off quickly as thermal inertia increases. For a thermal inertia of 160 tiu, the global mean change falls to 0.38\%, but remains about 1.5\% for facets facing the primary and a maximum of 9 K (3 \%). For a thermal inertia of 320 tiu, at the high end of the estimates for 1996 FG3, the mean change is 0.24\% (0.7 K) and maximum change is 1.9\% (6.1 K). Once thermal inertia reaches that of bare rock, or 2500 tiu, the maximum change has fallen to just below 1\%, or about 3 K.

\begin{figure}
    \centering
    \includegraphics[width=18cm]{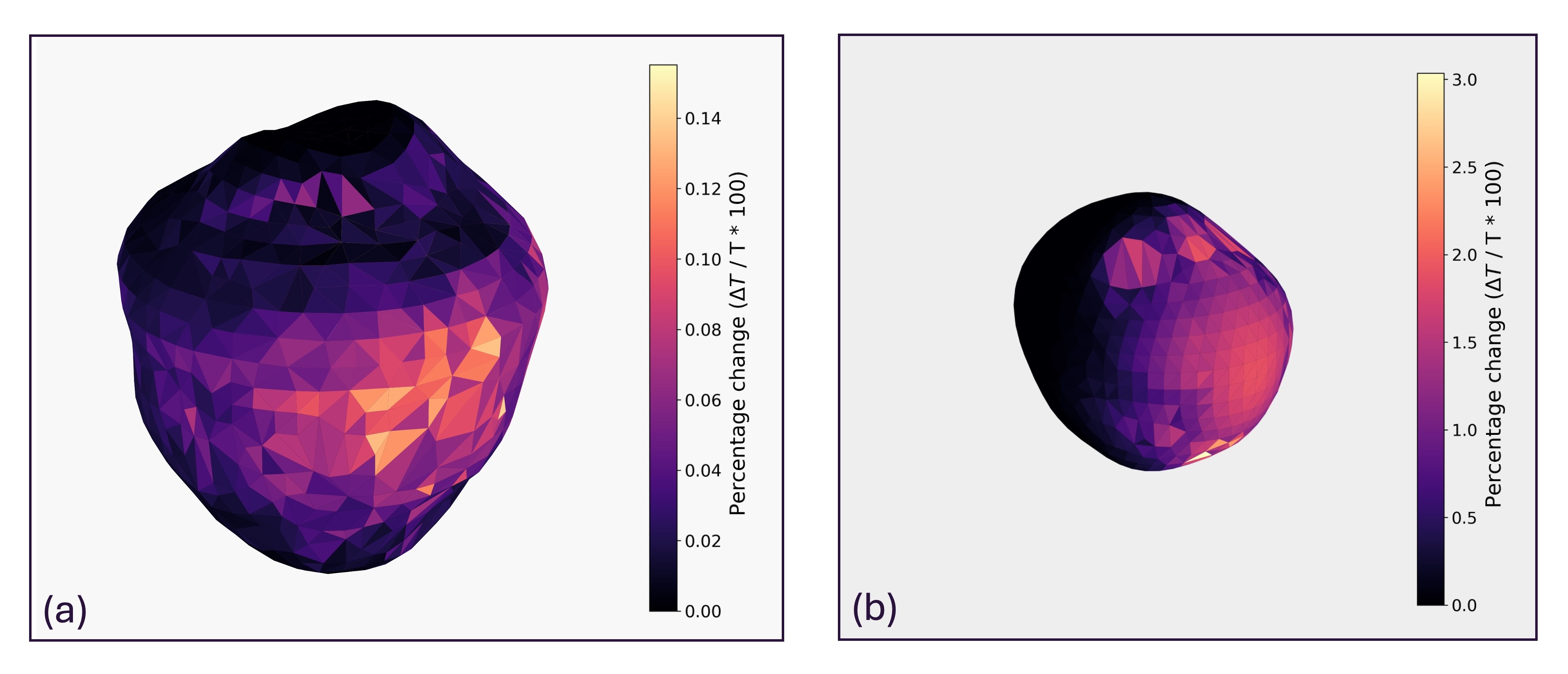}
    \caption{The maximum percent change in temperature caused by mutual radiation for each facet on the primary (a) and the secondary (b). Results shown on both bodies use a thermal inertia value of $\Gamma = 160 $ tiu. Note that the scale bars for each body are different, as the primary experiences about an order of magnitude smaller contribution from mutual radiation than the secondary. }
    \label{fig:moonshine}
\end{figure}

For the central thermal inertia case of $\Gamma = 160$ tiu for both bodies, the value for $M_{\mathrm{Max}}$ for each facet is overlaid on the shape models in Figure 
\ref{fig:moonshine}. Note the difference in scale bar for the two bodies. The primary receives significantly less mutual radiation exchange due to the secondary's smaller angular size. The map view highlights the influence of topography and orbit on mutual radiation. The inclination of each facet due to topography on each body can concentrate increased amounts of scattered radiation. On the secondary, the portion of the body that directly faces the primary shows a relatively smooth distribution of mutual radiation. However, the peripheral facets are less continuous, demonstrating a dependence on the topography of the secondary as well as primary topography seen by the secondary. 

On the primary, the variable radiation distribution is related to 1) the low resolution of the shape model 2) topography of the body and 3) the rotation of the body. As the primary rotates multiple times during one mutual orbit of the secondary, different facets experience different amounts of incident radiation from the secondary with each rotation. Figure \ref{fig:moonshine}, panel (a) shows one patch of facets that experienced locally higher mutual radiation. There is another patch on the opposite side of the primary. This distribution would be expected to equilibrate with model data from more mutual orbits. 

On the secondary, the maximum percentage change due to mutual radiation occurs just after the secondary has passed over the subsolar noon point on the primary. Hot daytime facets from the primary heat nighttime facets on the secondary. On the primary, the maximum percent change due to mutual radiation occurs on the facets that are exiting local night and are heated by secondary facets that have warmed after eclipse.  Both of these locations are pushed later as thermal inertia increases, consistent with the lag in peak temperatures expected from a higher thermal inertia.

\subsection{BYORP Results}
\begin{figure}
    \centering
    \includegraphics[width=15cm]{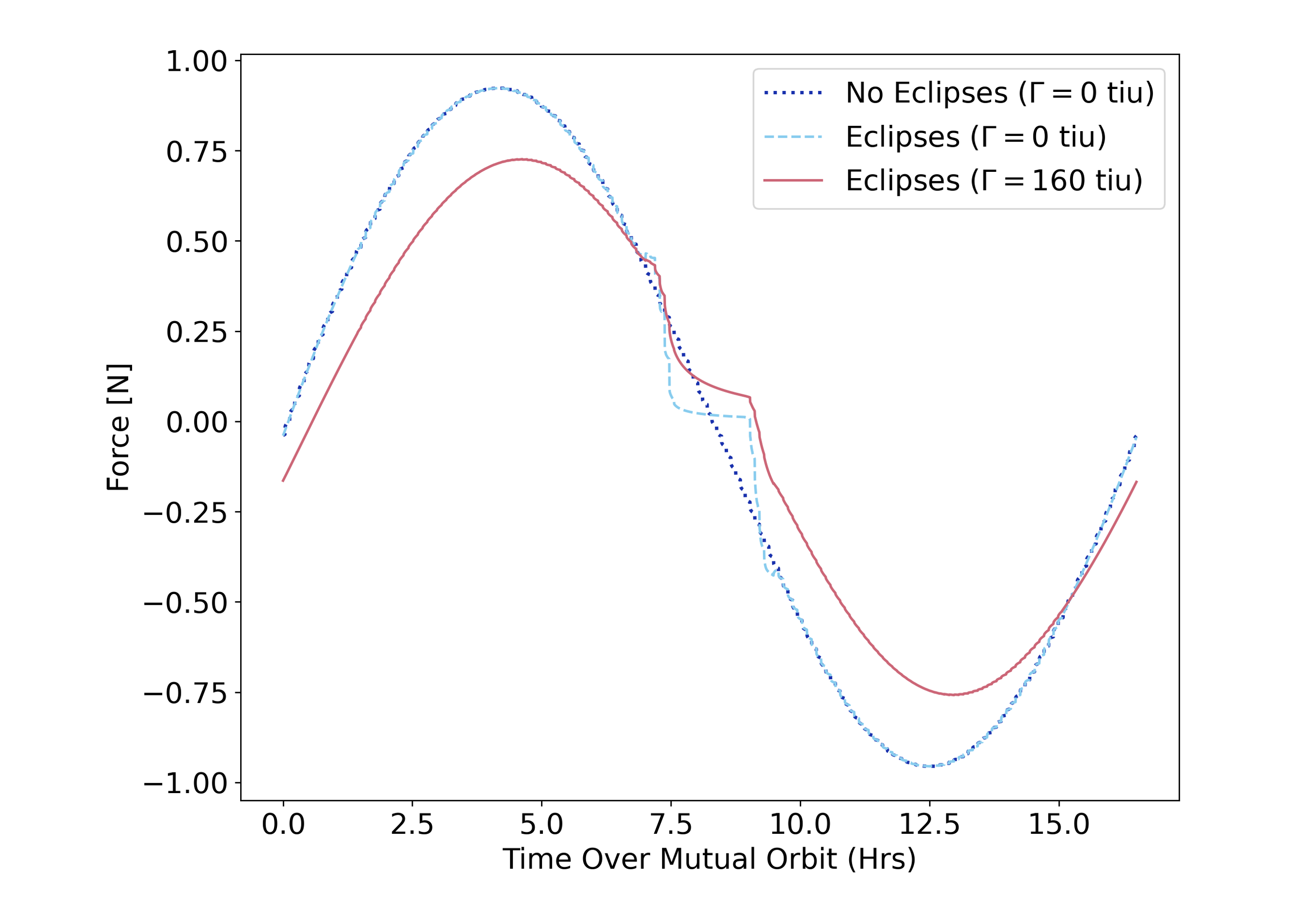}
    \caption{The force experienced by the secondary parallel to the direction of movement in the secondary fixed frame, or the y component of the force $F_{\mathrm{y}}$ for three model runs. The vertical axis is the force acting on the body in Newtons, and the horizontal axis is the time over which the secondary orbits the primary once. The dark blue dotted line shows the BTM run comparable to the theoretical case, including no eclipses and instantaneous temperatures. The dashed light blue line also uses instantaneous temperatures, but includes eclipses. The red line shows a BTM run with thermal inertia of 160 tiu and eclipses. These results are for a circular mutual orbit.}
    \label{fig:fyplot}
\end{figure}

Using the standard BYORP theory described in Section \ref{sec:BYORP}, we compute a validation value of $B = -0.0186960$. We used this analytical value to validate the model results from the BTM and to compare against the runs using thermal effects. Using the BTM with no eclipses, no mutual radiation and with instantaneous temperatures ($\Gamma = 0$), we are able to match this value to better than 0.01\%. A plot of the force parallel to the direction of movement in the secondary fixed frame, $F_{\mathrm{y}}$, is shown in Figure \ref{fig:fyplot}. The dark blue dotted line shows the case comparable to the theory. It is a smooth sinusoidal curve that begins and ends at 0. In contrast, the light blue dashed line representing a run with eclipses but no thermal inertia shows a deviation halfway through the mutual orbit, as the secondary experiences an eclipse. Finally, the solid red line shows a BTM run with eclipses and a realistic thermal inertia of 160 tiu. This line also deviates from the sinusoidal curve during the eclipse, but it is smoother due to the dampening effects of thermal inertia. Additionally, the entire curve is shifted slightly relative to the instantaneous temperature curves. This is due to the lag introduced by thermal inertia. Note that these results are for a circular mutual orbit. As we do not vary the distance between the bodies and assume no inclination of the secondary relative to the primary, this is representative.

\begin{figure}
    \centering
    \includegraphics[width=15cm]{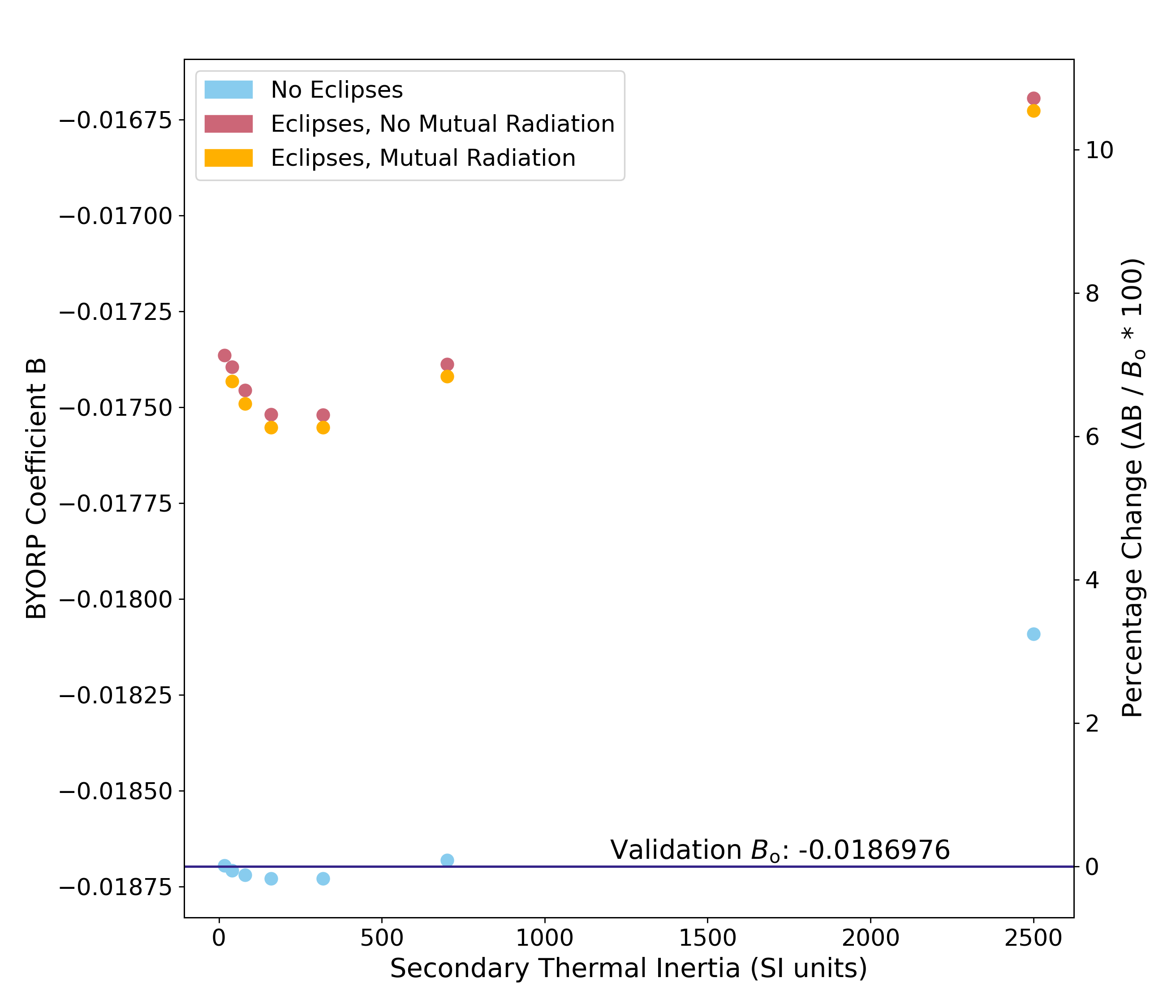}
    \caption{BYORP $B$ coefficients for various model runs plotted against the theoretical validation value $B_{\mathrm{o}}$ (shown with the solid line). The validation case uses no thermal inertia, no eclipses and no mutual radiation. Above the validation line indicates a smaller $B$, or smaller torque. Blue dots show BTM runs with no eclipses and no mutual radiation, but gradually increasing thermal inertia. Red dots show $B$ coefficients for BTM runs with eclipses and increasing thermal inertia, but no mutual radiation. Finally, the orange dots show BTM runs with all three effects. The vertical axis on the right side shows the percentage change relative to the validation $B_{\mathrm{o}}$, where a positive value indicates a less negative result, and therefore less torque.}
    \label{fig:byorpvals}
\end{figure}

BTM results for $B$ coefficients are shown in Figure \ref{fig:byorpvals}. The solid line represents the theoretical value $B_{\mathrm{o}}$ for zero thermal inertia and no mutual radiation exchange. A $B$ value above this line indicates less BYORP torque than suggested by the theory, while one below it indicates more BYORP torque. Several notable results stand out.

First, we note that the BTM is able to reproduce the validation case to a high degree of accuracy. Second, the largest deviation from the baseline value comes from the inclusion of eclipses. Holding the thermal inertia steady at 0 tiu and not including eclipses, the BYORP coefficient is altered by approximately 7\%, in the direction of less torque. Third, as thermal inertia increases, all three cases show a characteristic dip in $B$, indicating a larger BYORP value compared to a thermal inertia of 0. For no eclipses, this dip reaches a deviation of about -0.2\% relative to the theoretical value. For eclipses, the dip reaches a deviation of about 6.3\%. Between  $\Gamma = 160 $ and 320 tiu, this dip reverses, leading to a gradually less negative $B$ value. This dip is most evident in the range of thermal inertias estimated for 1996 FG3. As shown in the plot, as thermal inertia increases to 700 tiu (S-type asteroid-like) and 2500 tiu (bare rock-like), $B$ continues to move closer to 0. A bare rock secondary would be expected to have $B$ value approximately 3.2\% less negative than the theoretical value if no eclipses are present, and a 10.7\% difference if they are.

Interestingly, mutual radiation from the primary causes a nonzero but small effect, of order only one or two tenths of a percent change. It leads to a $B$ about -0.2\% more negative than the eclipsing value, although this difference decreases slowly as thermal inertia increases. 

We also held the secondary steady at $\Gamma = 160 $ tiu, and varied the primary's thermal inertia between 40 and 2500 tiu. However, these runs yield a $B$ value about 6.1\% less negative than the theoretical $B_{\mathrm{o}}$, the bulk of which is due to eclipsing. Changing the primary's thermal inertia leads to differences of only order 0.01\% between runs. They are thus not included in Figure \ref{fig:byorpvals} as they would appear as a single data point.

\section{Discussion}

In this work, we developed a 3D thermophysical model for binary asteroids. The model has immediate applicability to binary asteroid systems that are targets of mission flyby observations, as well as other airless bodies and planetary surfaces. It can be used to investigate characteristics of binary environments such as volatile stability, to generate exposure time estimates for cameras, and to interpret results from spacecraft infrared data. Using the BTM, we provide temperature modeling results for Janus target system 1996 FG3 at perihelion and for a range of estimated thermal inertias. These results, representing the hottest period of 1996 FG3's orbit, indicate the system can reach up to about 475 K during the day. The model also shows the strong temperature effect eclipses can have on a body. For a fully eclipsed secondary, some facets never experience a peak temperature at local noon, and instead experience large temperature drops. This leads to a strong asymmetry in temperature between the tidally locked sides of the body.

Our results also demonstrate the magnitude of the effects of mutual radiation on binary asteroid temperatures. For the 1996 FG3 system, the assumed thermal inertia has a large influence on the strength of radiation exchange between the primary and secondary. For a low thermal inertia of 40 tiu, the mutual radiation from the primary to the secondary can be large, with a mean contribution of well over 1\% and more than 10\% for the most strongly affected facets. These large increases are often ephemeral and vary throughout the mutual orbit. The magnitude of mutual radiation's contribution decreases quickly as thermal inertia increases. However, in the range of thermal inertias commonly found among small bodies, including Itokawa-like S-types with thermal inertias of roughly 700 tiu, there are still secondary facets that experience a maximum of more than a 1\% change in temperature due to mutual radiation. Thus, mutual radiation from primaries may be significant and need to be included for high resolution temperature modeling of binaries, especially at low thermal inertia. We also show that mutual radiation from the secondary to the primary consistently causes a mean difference in temperature of less than 1\% and can thus likely be excluded from thermal calculations. 

It is important to note that the magnitude of the effects of eclipses and mutual radiation will vary according to the individual properties of the system. For more widely separated or highly inclined binary systems, eclipsing may not occur and mutual radiation may be insignificant even at low thermal inertia. However, for closely orbiting or large secondaries, these effects may be exacerbated and cause a comparably larger effect than that described here. 

Additionally, we expand this model to estimate the magnitude of the binary YORP effect. This is useful both for testing the assumptions of the current theory and informing what properties affect binary dynamics. We validate this model against the current BYORP theory described in \cite{McMahon2010SecularBYORP} based off the methodology described in \cite{Scheeres2007TheYORP}, and match the theoretical value to a high level of precision. From these results, we observe several interesting trends. 

First, eclipses cause a deviation of order several percent in the BYORP coefficient $B$, making $B$ less negative, and thus indicating a reduction of torque. By reducing temperatures on the body, they also reduce the temperature amplitude. This in turn decreases the force generated by emitted infrared photons. The secondary facets most affected by eclipses are those facing the primary. However, the decrease in emission from these facets will mostly be oriented along the line between the primary and the secondary, or $F_{x}$, which does not affect BYORP. The deviation of $B$ values produced by the BTM arises from the shadowing of facets on the side. Reducing the emission in the positive and negative along track direction causes a dip in the $F_{\mathrm{y}}$ force, and thus leads to a less negative $B$.

\begin{figure}
    \centering
    \includegraphics[width=15cm]{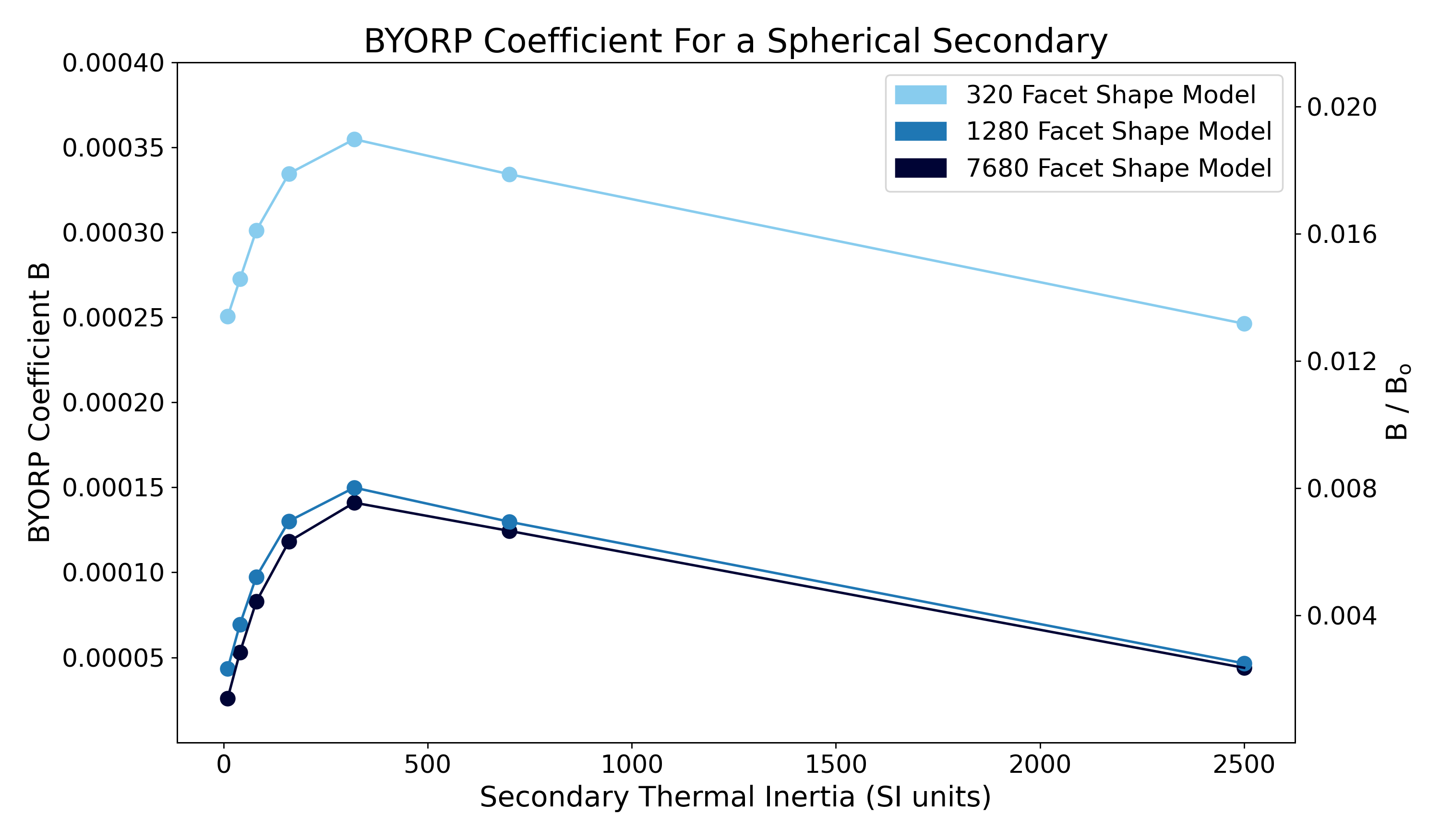}
    \caption{BYORP $B$ Coefficients for a spherical secondary. The data points shown do not include eclipses. The three colors denote different resolutions of spherical shape model. Note that the $B$ values are over two orders of magnitude smaller than the normal BTM runs. The axis on the right highlights this by showing the calculated spherical $B$ values when compared against the validation $B_{\mathrm{o}}$ for 1996 FG3.}
    \label{fig:byorpsphere}
\end{figure}

The second trend observed arises from thermal inertia. As thermal inertia increases, it initially causes an increase in torque, or more negative $B$, before beginning to move towards 0. This dip is demonstrated by each group of data in Figure \ref{fig:byorpvals}. This implies that at low $\Gamma$, the delay in when peak temperatures occur on the body caused by thermal inertia introduces some asymmetry to the force, but diurnal amplitudes are still high. As the thermal inertia increases, it serves to dampen the amplitude of surface temperatures, effectively working towards an equilibrium across all sides of the body. A value of thermal inertia between $\Gamma = 160$ and $\Gamma = 320$ tiu appears to be a tipping point where the slight torque induced by the lag from thermal inertia is balanced by the damping effect on temperatures. Together with eclipses, a high thermal inertia of 2500 tiu leads to the maximum deviation from the theoretical value, but the least amount of torque on the secondary's orbit. Previous work has assumed that BYORP should be independent of thermal inertia \citep{McMahon2010DetailedPopulation, Cuk2005EffectsNEAs}. However, we find based on these results that thermal inertia does have a small but nonzero effect, especially when combined with eclipses.

To determine whether the asteroid shape plays a role in the shape of the thermal inertia curve, we ran additional BTM runs where the secondary was replaced with spheres of varying resolution (320, 1280 and 7680 facets). The results for the spherical models, with no eclipses but with increasing thermal inertia values, are shown in Figure \ref{fig:byorpsphere}. For the theoretical case, a spherical object with no eclipses and instantaneous temperatures should have a BYORP coefficient of $B = 0$. We match this data point to better than 0.0001. However, as $\Gamma$ increases, $B$ initially grows to be a positive value, as opposed to the negative value for 1996 FG3, and then falls back as the high thermal inertia dampens temperatures. This indicates that the shape of the thermal inertia curve is dependent on asteroid topography, rather than solely on the value of $\Gamma$. It is important to note that all values for the sphere are at least 100 to 1000 times smaller than the coefficients for 1996 FG3, as expected from a surface with no topography. The division of the spherical shape into triangular facets means the sphere is imperfect, which may lead to very small but nonzero $B$ values. To test this effect, we ran the same suite of models using a low resolution sphere (320 facets) and a higher resolution sphere (7680 facets), in addition to the 1280 facet sphere. Though the $B$ values output by the model were all small compared to those obtained for 1996 FG3, the lower resolution sphere led to higher $B$ values, and the higher resolution sphere decreased the magnitude of $B$ values. These results demonstrate that even the high resolution sphere experiences a very small amount of BYORP. This is likely the result of numerical error. The small nonzero value of $B$ for the zero thermal inertia case, about 0.00002, couples to the thermal effects, producing the curve shown.  

We find eclipsing, mutual radiation and thermal inertia all change $B$ by a nonzero amount. For all of these thermal effects, asteroid shape and orbital configuration will influence the strength of the deviation they cause. Thus shape of the secondary can be considered a first order effect, eclipsing and thermal inertia a second order effect, and mutual radiation is a third order effect with a minimal but nonzero change to the BYORP coefficient.  

These effects are small, with most runs with realistic binary effects and thermal inertia causing a deviation to $B$ of a few percent. However, over many orbits, this small perturbation may cause long term changes to a binary's orbit. A detailed consideration of how the $\Delta B$ demonstrated here affects long term orbital dynamics will be the subject of further study. However, the magnitude of this perturbation can be estimated by calculating the change in semimajor axis and eccentricity of the binary mutual orbit. \cite{Jacobson2011LONG-TERMASTEROIDS} provides equations from the first order theory for the evolution of these two parameters, adapted here: 

\begin{equation}
    \dot{a}_{B} = \pm \frac{3 H_\odot B}{2 \pi} \left( \frac{\tilde{a}^{3/2}}{\omega_{d} \rho R^{2}_{p}} \right) \frac{\sqrt{1 + q}}{q^{1/3}} R_{\mathrm{p}}
    \label{eq:aDot}
\end{equation}

\begin{equation}
    \dot{e}_{B} = \mp \frac{3 H_\odot B }{8 \pi} \left( \frac{\tilde{a}^{1/2} e}{\omega_{d} \rho R^{2}_{p}} \right) \frac{\sqrt{1 + q}}{q^{1/3}}
    \label{eq:eDot}
\end{equation}

where $H_\odot = F_\odot / (a^{2}_{\odot} \sqrt{1 - e^{2}_{\odot}})$, and $\tilde{a} = a/R_{\mathrm{p}}$ is the normalized semimajor axis. Here $F_\odot$ is the solar radiation constant, approximately equal to $1\times 10^{14}$ kg km $\mathrm{s}^{-2}$ \citep{McMahon2010DetailedPopulation}, $a_\odot$ and $e_\odot$ are the heliocentric semimajor axis and eccentricity, $q$ is the ratio of the secondary mass to the primary mass and $R$ is the radius of the primary. $\omega_{d} = (4 \pi G \rho / 3)^{1/2}$ is the surface disruption spin limit for a sphere, and $G$ is the gravitational constant. Note that the BYORP effect can shrink or expand the semimajor axis of the binary, but will have an effect on eccentricity opposite in sign to that of $a$. 

Using these relationships, we estimate the effect of the changes to the BYORP coefficient arising from eclipses, the thermal effect that causes the largest $\Delta B$. To consider eccentricity, we use $e = 0.07$, the maximum possible eccentricity for the 1996 FG3 mutual orbit from \cite{Scheirich2015TheEvolution}. Equations \ref{eq:aDot} and \ref{eq:eDot} indicate that for the validation $B_{\mathrm{o}}$ value, the semimajor axis would evolve at a rate of approximately $-8.8 \times 10^{-10}$ m/s, and the eccentricity would increase by about $6.2 \times 10^{-15}$ $\mathrm{s}^{-1}$. When compared to the BYORP validation $B_{\mathrm{o}}$, the introduction of eclipses results in a roughly 7\% difference in $B$, and a comparably lower torque. This leads to a smaller rate of change of the semimajor axis ($-8.2 \times 10^{-10}$ m/s) and eccentricity ($5.8 \times 10^{-15}$ $ \mathrm{ s}^{-1}$). Over 1000 years and in the absence of tidal forces, this implies that eclipses could reduce the contraction of the semimajor axis by approximately 2 meters. It also implies an increase in eccentricity, albeit a smaller one than the validation value by  $\sim 1.4 \times 10^{-5}$. Over 10,000 years, this difference in semimajor axis grows to about 20 meters, and difference in eccentricity to $\sim 10^{-4}$. Though this difference is small, previous work has demonstrated that eccentricity evolution has a marked effect on overall system evolution. Increasing the eccentricity of a binary mutual orbit can excite other dynamical effects, such as libration. Depending on the system, these effects may be damped by tides, or the libration may grow and result in an attitude instability \citep{Cueva2024ThePost-DART}.  

\cite{Scheirich2015TheEvolution} provide a measurement of the orbital evolution rate for 1996 FG3 of approximately $-0.07 \pm 0.34$ cm/yr. When converted to comparable units, the drift arising solely from BYORP, and specifically from the validation value $B_{\mathrm{o}}$, leads to a theoretical drift of approximately -2.78 cm/yr. The difference between the measured drift and theoretical value is thought to arise from the effects of tidal forces, which work to mitigate the effects of BYORP. Thus the measurement from \cite{Scheirich2015TheEvolution} cannot be directly compared to the values determined by this work. Nonetheless, this demonstrates the variety of factors at play in asteroid dynamics. 

One critical factor is shape. Asteroid shape models are often low resolution, with only a few hundred or thousand facets. This uncertainty in shape can affect the predicted value of BYORP. The true value of the $B$ coefficient for the 1996 FG3 system is likely dependent on features unresolved by the current shape model. This is supported by recent work by \cite{Roberts2021RotationalStrength}, which found that both large and small scale topography are potentially significant for YORP spin-up. Changes in shape to the 1996 FG3 secondary are likely to affect $B$, and could cause $B$ to increase or decrease. However, it is unlikely to have a larger effect on $B$ than that from large scale topography.

Multiple avenues for further study exist using the BTM, such as the inclusion of roughness in temperature calculations. Recent studies have suggested that YORP values are sensitive to changes in roughness (e.g., \cite{Statler2009ExtremeTopography}), and the same would be expected for BYORP. Additionally, the effect of secondary shape, including body axis ratios, as well as orbital factors such as separation distance and libration could be included in future versions of the model. 
 
Measuring the BYORP value may help constrain information about how energy dissipates through the body, offering insights into binary asteroid interiors \citep{Quillen2022Non-principalEffect}. Energy arising from BYORP torque can be dissipated via tides, an interaction often described through the tidal Love number, $k_{2}$, and the tidal quality factor, $Q$ \citep{Goldreich2009TIDALPILES,Jacobson2011LONG-TERMASTEROIDS}. For binary systems that may be in tidal-BYORP equilibrium, such as 1996 FG3, accurately modeling BYORP may give insight into the tidal strength and the parameters $Q$ and $k_{2}$ \citep{Cueva2024ThePost-DART}. Recent work by \cite{Zhou2024TheAsteroids} also found that the binary Yarkovsky effect could be a powerful mechanism for synchronizing secondaries, and may be especially important for highly separated systems. In order to distinguish between the effects of tides, BYORP and binary Yarkovsky, it is critical to understand what effects may impact the theoretical value. The implications of thermal effects on the determination of asteroid interior properties is beyond the scope of this study, and should be investigated in future work.

\section{Conclusions}
This work contains a description of a new thermophysical model for binary asteroids, as well as a new analysis and tool to study the Binary YORP effect. We summarize our key findings and current results as follows: 

\begin{enumerate}
    \item We describe the Binary Thermophysical Model, or BTM, a 3D thermophysical model for interacting binary pairs.
    \item We present temperature estimates for binary system (175706) 1996 FG3 for a range of thermal inertia values. 
    \item For 1996 FG3, mutual radiation can alter surface temperatures of the secondary for short periods. For a low thermal inertia of 40 tiu, the mean maximum change in temperature for secondary facets facing the primary is over 1\%.
    \item We present a new method for estimating the BYORP coefficient $B$ for binary systems using temperatures and forces produced by the BTM. Using this model, we successfully replicate the existing value for 1996 FG3's $B_{\mathrm{o}}$ from pure BYORP theory.
    \item With this model, we include second-order thermal effects such as eclipsing, thermal inertia and mutual radiation, and we present how they alter $B$. 
    \item Eclipses are the most powerful second-order effect. We find they alter the value of $B$ for the 1996 FG3 system by approximately 7\% without thermal inertia. When thermal inertia is included, this change ranges from about 6\% to over 10\% for very high thermal inertia values.
    \item We find that, counter to previous work, thermal inertia has a small but nonzero effect on $B$, especially when paired with eclipses. 
    \item Mutual radiation from the primary causes a nonzero but minimal change to $B$. For 1996 FG3, this effect is less than 1\%. 
    \item For 1996 FG3, thermal effects serve to reduce the torque on the binary relative to that predicted by existing BYORP theory. In the absence of tidal effects, this change to $B$ could result in reduced contraction of the semimajor axis. 
\end{enumerate}

Though the models presented here are particular to the 1996 FG3 system, our results indicate that thermal effects may need to be considered when calculating $B$ for binary asteroid systems. Eclipses and thermal inertia are both demonstrated to affect the BYORP coefficient. Certain effects, such as mutual radiation, may not prove significant for some systems and can thus be ignored, but may play a larger role for close in systems with large primaries. Thus, the relative importance of each effect on the BYORP coefficient for a system will vary for each binary being studied. 

\section{Acknowledgments}
\begin{acknowledgments}
The authors thank and acknowledge the Janus science team for their thoughtful commentary and contributions. This work was supported by the Janus mission and contributions by K. C. S. and R. H. C. were both partially supported by the National Science Foundation Graduate Research Fellowship under Grant No. DGE 2040434. Any opinions, findings, and conclusions or recommendations expressed in this material are those of the author(s) and do not necessarily reflect the views of the National Science Foundation.

\end{acknowledgments}

\bibliographystyle{plainnat} 
\bibliography{references}{}


\end{document}